\documentclass[12pt]{article}

\usepackage{amsmath,amsfonts,amssymb,latexsym,graphicx}

\numberwithin{equation}{section}
\def\be{\begin{equation}}
\def\ee{\end{equation}}
\def\bq{\begin{eqnarray}}
\def\eq{\end{eqnarray}}
\def\beq{\begin{eqnarray}}
\def\eeq{\end{eqnarray}}

\def\a{\alpha}
\def\b{\beta}

\def\t{\theta}

\def\pa{\partial}
\def\bs{\boldsymbol}

\begin{document}
\title{\textsc{Onset of synchronization in coupled Mixmaster oscillators}}
\author{\Large{\textsc{Spiros Cotsakis}\thanks{skot@aegean.gr}}\\
Institute of Gravitation and Cosmology, RUDN University\\
ul. Miklukho-Maklaya 6, Moscow 117198, Russia\\ and\\
Research Laboratory of Geometry,  Dynamical Systems  and Cosmology\\
University of the Aegean, Karlovassi 83200, Samos, Greece}
\date{July, 2021}
\maketitle
\newpage
\begin{abstract}
\noindent We consider the problem of asymptotic synchronization of different spatial points coupled to each other in inhomogeneous spacetime and undergoing chaotic Mixmaster oscillations towards the singularity.
We demonstrate that for couplings larger than some threshold value,  two Mixmaster spatial points $A,B$, with $A$ in the past of $B$, synchronize and thereby proceed in perfect unison towards the initial singularity.
We further show that there is a Lyapunov function for the synchronization dynamics that makes different spatial points able to synchronize exponentially fast in the past direction.
We provide an elementary proof of how an arbitrary spatial point responds to the mean field created by the oscillators, leading to their direct interaction through spontaneous synchronization.
These results ascribe a clear physical meaning of early-time synchronization  leading to a resetting effect for the two BKL maps corresponding to  two distinct oscillating spatial points, as the two maps converge to each other to  become indistinguishable at the end of synchronization. Our results imply that the universe generically organizes itself  through  simpler, synchronized, states as it approaches the initial singularity.
A discussion of further implications of early-time inhomogeneous Mixmaster synchronization is also provided.
\end{abstract}
\newpage
\tableofcontents
\newpage
\section{Introduction}
Ever since the appearance of the  theorems about the existence of spacetime singularities in the sense of geodesic incompleteness of generic solutions in general relativity and cosmology \cite{hp}, \cite{he}, and  the first indications that the general solution cannot have a singularity of a power-law character  \cite{bkl1}, the true nature of generic inhomogeneous solutions of the full Einstein equations towards the cosmological singularity has been a formidable challenge. In fact, the initial duality between the proved global geodesic incompleteness and the expected possibility of local divergences in geometric and physical invariants, has been a major part of the cosmological singularity problem ever since. It is still ill-understood, but one that  has  stimulated a lot of  research.

For the cosmological singularity problem  many conjectures have been formulated. One such conjecture is about the  asymptotic behaviour of the solutions to the gravitational equations, and is based on subsequent work of V. A. Belinski, I. M. Khalatnikov, and E. M. Lifshitz (hereafter abbreviated to `BKL'), the BKL conjecture \cite{bkl2,bkl3}.
This `BKL singularity' is essentially of a new and eminently complicated type, and no power-law behaviour may be ascribed to it. In the special, homogeneous case, it is based on a generalized Kasner solution and a subsequent `Bianchi IX' oscillatory behaviour  having an endless sequence of Kasner epochs grouped into eras all the way to the finite time past singularity, cf. \cite{bkl2}. However,  in the general, inhomogeneous case,  each spatial point although evolving like one of these separate Bianchi IX homogeneous universes,   experiences the collective gravitational perturbations of neighboring spatial points. The BKL conjecture \cite{bkl3} then  describes the generic behaviour  by  roughly stating that almost all solutions of the Einstein equations approach an initial spacelike, vacuum-dominated, local, and oscillatory singularity, a regime which eventually becomes stochastic and essentially random \cite{chaos5,chaos6}.

The  BKL conjecture is essentially an asymptotic statement about the behaviour of generic inhomogeneous spacetimes.   To a leading-order approximation, the generic inhomogeneous solution in the past direction will consist of an infinite series of (generalized) Kasner states that alternate as in the exact `Mixmaster' solution first discovered in \cite{bkl2},\cite{misner}, but where now the defining constants (four in  vacuum) become arbitrary \emph{functions} of the three spatial coordinates. Early investigations showed that the behaviour of small inhomogeneous perturbations to the original Mixmaster evolution sustain that picture (cf. \cite{hu1}-\cite{hu2}, and references therein), but later it was claimed that other inhomogeneous instabilities and/or gravitational wave and other kinds of perturbations might arise, leading to  difficulties in accepting this picture (in technical language, we refer here to the property of `linearization instability' and the existence of spurious perturbations \cite{li1}-\cite{li5}, as well as the existence of  `spikes' \cite{hul}).

There are, however, precise mathematical results about the behaviour of spatially homogeneous $G_3$ solutions \cite{bogo,we,ring}, and about more general, inhomogeneous ones, the so-called $G_2$  solutions \cite{euw}, which can also be given a description in terms of a past attractor (see also \cite{ycb09}, \cite{dabu} for recent Hamiltonian methods). However,  for the general $G_0$ case without any symmetries,  only conjectures can be framed, cf. \cite{uvwe} for a precise presentation (see also Refs. \cite{berger1}-\cite{berger5} for other details and numerical  results). In this paper, we shall follow the notation of Ref.  \cite{uvwe}.

But perhaps a major obstacle  for the plausibility and proof of the BKL conjecture is the fact that  the behaviour of the homogeneous Mixmaster universe is already  formally a chaotic system \cite{chaos1}, \cite{chaos2}. It possesses a smooth invariant measure \cite{chaos3}, with a corresponding shift map \cite{chaos4}, and a detailed statistical theory for the Mixmaster stochastic behaviour can be readily built \cite{chaos5}, \cite{chaos6}. Another complication, possibly related to the chaotic behaviour, this time  seen in the generic evolution of each spatial point in the inhomogeneous case, is connected with the so-called gravitational turbulence \cite{bel}. This is also related to the so-called process of  fragmentation, or cellular structure, an effect  discovered by Kirillov and Kochnev \cite{kir1,kir2}, and independently by Montani  \cite{mon}, also recently elaborated by Barrow \cite{bafra}. This process, as described in these references, is supported by a claimed monotonic increase  of spatial gradients of the metric on approach to the singularity. It arises through the transport of energy from large to small scales, because  the Mixmaster oscillatory regime in the inhomogeneous case is `unstable' and may decay into oscillations of smaller scales. This then may produce a phenomenon similar to 3-dimensional turbulence in fluid mechanics, but in this case the transport may continue down to zero scale.

The emerging picture of infinitely many coupled spatial points oscillating chaotically and indefinitely towards the past singularity, exhibiting chaos on all scales in a spatio-temporal manner,  makes it plausible to suppose that it may soon get unstable and continue as an out-of-control system exhibiting complex behaviour.  This is perhaps related to the acknowledged difficulty in obtaining universally acceptable results with clear descriptions and explanations. Indeed, even for the homogeneous Mixmaster case, there are few completely established facts, cf. \cite{heinzle}.

The problem  of understanding the behaviour of a complex system that includes an arbitrary number of (often chaotically) oscillating  subsystems coupled to each other in different ways is, of course, a central one in modern nonlinear dynamics. This problem is usually characterized, on the one hand, by the inexorable increase of entropy towards states of ever greater disorder, and on the other, by the ability of subsystems to assemble together and form diverse structures we often see all around us. This enigma poses a profound mystery, that of the appearance of order, not only  in space but also \emph{in time}-the more subtle property called \emph{synchronization} (commonly called `sync', a name we  adopt below). When do coupled oscillators synchronize themselves, how long can sync last, when or how can it break down, and what are the implications of sync for the fate of the system?

Sync is truly ubiquitous in science, and despite intensive work that has been done over the past few decades, it is still not well-understood (see \cite{stro03} for an exceptionally clear popular account of the diverse aspects of sync in nature). Remarkably,  attempts to describe the breath-taking manifestations and implications of sync even extend to chaotic systems oscillating erratically. Chaotic synchronization is a well-known phenomenon in which one employs advanced dynamical techniques to successfully link two, or more, coupled autonomous chaotic (sub-)systems that map out the same attractor in phase space, but their trajectories  become uncorrelated in time. By `linking' we mean roughly that their phase curves converge  to each other, a property which appears to be structurally stable, and the coupling goes to zero so that the systems behave as the underline uncoupled system \cite{pc}, \cite{elp}.

Perhaps somewhat unexpectedly, sync is not well-known in cosmology, but very recently  Barrow studied a specific coupling of two different Mixmaster regions undergoing chaotic oscillations in a generic inhomogeneous spacetime. He concluded that the coupled regions may join together and `synchronize', so that a simplified past evolution consistent with the homogeneous Mixmaster picture as well as the BKL conjecture may result \cite{ba20}. If this effect is true, then some of the difficulties that arise in attempts to describe part of the past evolution of the general solution of the Einstein equations towards the initial singularity may be  alleviated, and the plausibility of the BKL conjecture be accordingly enhanced.

It is with the purpose of giving a feeling of the possible importance and wider implications of sync to cosmology that  we introduce and study the possibility of sync between oscillating spatial points in the inhomogeneous Mixmaster setup in a somewhat systematic way.

The plan of this paper is as follows.
In the next Section, we establish our notation and,  after certain preliminary developments, arrive at the dynamical equations that describe the evolution of spatial points in inhomogeneous Mixmaster spacetime.
In Section 3, we introduce  basic concepts of synchronization in general nonlinear dynamics, and  set up the Mixmaster sync problem by splitting the expansion-normalized variables into driving and responding ones, and introducing the synchronization function $\Omega$ between two Mixmaster spatial point trajectories undergoing oscillations.  We also provide an analysis of Mixmaster synchronization using Lyapunov exponents corresponding to individual Mixmaster spatial trajectories, and show how they are intimately connected with the ability of the inhomogeneous spatial points to synchronize after some critical threshold of their coupling is passed.
In Section 4, we prove the existence of a Lyapunov function in the synchronization dynamics, for different magnitudes of the  shear variables of distinct  spatial points, first for the case of zero coupling, and then for general coupling of the spatial points.
In Section 5, we study phase synchronization of the dynamics using the Barrow-Kuramoto coupling as an example,  provide an elementary proof of the dynamical equation describing the coupling of one Mixmaster spatial point to the mean field, and show how this  leads to a direct coupling between two spatial points.
In Section 6, we study the implications of sync to the possibility of eventual convergence to a common sequence of values of two BKL maps that correspond to two different spatial points, and discuss the resulting physical interpretation of the effects of sync.
We conclude in the last Section with a discussion of other issues not considered in  the present work, as well as further conjectural implications that follow directly from our results.

\section{Evolution of spatial Mixmaster points}
It is a basic fact that in the  inhomogeneous Mixmaster model \emph{each spatial point evolves like a separate homogeneous Mixmaster universe} as we approach the initial singularity. This characteristic property of the model plays a central role in BKL and in more recent approaches to the problem,  and leads to formulations of the Einstein field equations as systems of partial differential equations where conjectures about the properties of the solutions have been formulated. In this paper, we follow a simpler approach that  focuses instead on the behaviour of the interaction of two neighboring spatial points on approach to the singularity.

\subsection{Hubble-normalized state}
Our ensuing analysis applies to the $G_0$ cosmologies that admit no isometries and have a state that is described in vacuum by the Hubble-normalized state vector \cite{uvwe},
\be\label{sv}
\mathbf{X}=(E_\a^i,r_\a,\Sigma_{\a\b},N_{\a\b},A^\a)^{T}.
\ee
Explicitly\footnote{Latin indices run from 0 to 3, and Greek indices from 1 to 3.}, we introduce a fundamental 4-velocity timelike vector field $\textbf{u}$ on the cosmological spacetime $(M,g)$, and use the simple fact that  the coordinate basis $(\pa_t,\pa_\a)$ is not orthogonal on a local chart $(U,\phi)$ of the manifold $M$, because the $\pa_i$'s are inverse images of the canonical basis vector fields of $\mathbb{R}^4$ under a $\phi$-induced map on each tangent space.

Then in a cosmological setting one prefers instead to use an \emph{orthonormal frame} $(\mathbf{e}_0,\mathbf{e}_\a),$ which singles out a future-directed, unit, timelike vector field  $\mathbf{e}_0$, usually equal to $\textbf{u}$, supplemented by an orthonormal spatial frame $(\mathbf{e}_\a),$ in the rest 3-spaces of $\mathbf{e}_0$, making the frame metric the Minkowski metric $\eta_{ab}=\textrm{diag}(-1,1,1,1)$. It is simply related to the coordinate frame  by the equations
\be
\mathbf{e}_0=N^{-1}(\pa_t-N^\a\pa_\a),\quad  \mathbf{e}_\a=e_\a^\b\pa_\b,
\ee
with $N,N^\a$ being the lapse and shift functions.

This orthonormal frame is further used to introduce the \emph{Hubble-normalized frame} $(\boldsymbol{\pa}_0,\bs{\partial}_\a)$ with,
\be
\bs{\pa}_0=\frac{1}{H}\mathbf{e}_0,\quad\bs{\pa}_\a=\frac{1}{H}\mathbf{e}_\a \equiv E_\a^\b\pa_\b,
\ee
and new Hubble-normalized variables defined as follows:
\be
(\dot{U}^\a,\Sigma_{\a\b},A^\a,N_{\a\b},R^\a)=\frac{1}{H}(\dot{u}^\a,\sigma_{\a\b},a^\a,n_{\a\b},\Omega^\a).
\ee
Here the Hubble scalar function $H$ is defined to be $1/3\Theta,$ where the expansion scalar $\Theta$ is  $\Theta=\nabla_a u^a$. Also we introduce the acceleration vector field $\dot{u}_\a={u}_\b\nabla^\b u_\a$, giving the time derivative of  $\textbf{u}$  as a directional derivative along $\textbf{u}$ (also same notation for the time derivative of any tensor, e.g.,  $\dot{\rho}_{ab}=u^c\nabla_c \rho_{ab}$). In addition, $\sigma_{\a\b}$ denotes the shear rate tensor which depends on $\textbf{u},\Theta,\dot{u}_\a$, and the projection tensor is defined by $h_{ab}=g_{ab}+u_au_b$. The symbols $a^\a,n_{\a\b}$ determine the connection of the spacelike hypersurfaces $\Sigma$ normal to the fundamental vector field $\textbf{u}$, and $\Omega^\a$ is the local angular velocity of the spatial frame  $(\mathbf{e}_\a)$. One can further show that the state vector field $X(t,x^\a)$ given by Eq.  (\ref{sv}) satisfies a system of evolution equations with constraints (cf. \cite{uvwe} and refs. therein), the specific form of which is however not necessary for our present purposes.

\subsection{Variables along orbits of spatial points}
In what follows, we are interested in the evolution of individual spatial points, corresponding to the evolution of  the Hubble-normalized vector field $X(t,x^\a)$ for \emph{fixed} $x^\a$. Each such fixed spatial point describes a different orbit in a \emph{finite-dimensional} state space, and we shall be concerned with the direction of approach of all these orbits to the past singularity as $t\rightarrow 0^+$.

According to the BKL scenario and subsequent analyses, each such orbit corresponds to the dynamics of an individual homogeneous Mixmaster universe (Bianchi type IX). In this case, the  Hubble-normalized variables and state vector field $X$ simplify because along each orbit of a spatial point $A$ in an inhomogeneous Mixmaster spacetime, we have
\be
\bs{\pa}_\a (\dots)=0,\quad \dot{u}^\a=0,\quad\textrm{along an orbit}.
\ee
This implies that the spatial operators $\bs{\pa}_\a$ do not appear in the equations taken along such orbits. Also all variables along any such orbit become functions of only the time $t$, because their derivatives with respect to the spatial variables are zero. In particular, along the orbit of a spatial point the only non-zero commutation   functions are $H,\sigma_{\a\a}, n_{\a\a}$, which we can write in the usual form,
\be
\sigma_{\a\b}=\textrm{diag}(\sigma_{11},\sigma_{22},\sigma_{33}) ,\quad n_{\a\a}= \textrm{diag}(n_1,n_2,n_3).
\ee
Since $\sigma_{\a\b}$ is traceless, we can label its two independent components using the variables,
\be
\sigma_+=\frac{1}{2}(\sigma_{22}+\sigma_{33}),\quad \sigma_-=\frac{1}{2\sqrt{3}}(\sigma_{22}-\sigma_{33}),
\ee
so that the dimensionless state $X$ along the orbit of a fixed spatial point $A$ becomes,
\be
X_A (t)=(\Sigma_+,\Sigma_-,N_1,N_2,N_3),\quad\Sigma_\pm=\frac{\sigma_\pm}{H},\,\,N_\a=\frac{n_\a}{H}.
\ee
In the following, we shall also take the time derivatives with respect to the dimensionless $\tau$-time related to the proper (clock) time by,
\be
\frac{dt}{d\tau}=\frac{1}{H}.
\ee
Then the interpretation of the variables in the dimensionless state vector field $X_A (t):=X(\,\cdot\, ,x^\a (A))$, which describes the orbit of a spatial point $A$ in inhomogeneous spacetime, is completely analogous to that in the usual Bianchi cosmology, cf. \cite{we}, chap. 6, with the important difference that in the latter case there is only one such orbit-the whole universe is homogeneous. Therefore $\Sigma_\pm$ describe the anisotropy in the Hubble flow, and $N_\a$  describe the spatial curvature and the Bianchi type of the isometry group \emph{during the evolution along the orbit of the spatial point $A$}.

\subsection{Relation to the BKL and Hamiltonian variables}
The relation of the dimensionless variables $(\Sigma,N)$ to the Hamiltonian (Misner) and the  BKL variables, which we briefly discuss here, is the standard  one   described in many references, eg.,  \cite{we}, chapters 10, 11.

To compare with the Misner variables, we introduce a new orthonormal `Misner'-frame $(\mathbf{n},\mathbf{E}_\a)$, with $\mathbf{n}$ unit and orthogonal to the group orbits $t=\textrm{const.}$ (eg., $\pa_a t$) and tangent to a geodesic congruence, and $\mathbf{E}_\a$ being spatial, time-independent vector fields, i.e., $[\mathbf{n},\mathbf{E}_\a]=0$, leading to the line element $ds^2=-dt^2+g_{\a\b}(t)\mathbf{W}^\a \mathbf{W}^\b$, where the 1-forms $\mathbf{W}^\a$ are dual to $\mathbf{E}_\a$. Then, introducing the lapse by $dt=N(\tilde{t})d\tilde{t}$, the orthonormal frame $(\mathbf{e}_0,\mathbf{e}_\a)$ introduced above is related to the Misner frame by
\be
\mathbf{e}_0=\frac{1}{N}\frac{\pa}{\pa\tilde{t}},\quad\mathbf{e}_\a=e^{-\b_\a}\mathbf{E}_\a,
\ee
where $\b_\a$ are the standard Misner variables, or, in terms of the Hamiltonian variables $(\b^A,p_A),\,\,(A=0,+,-),$ we have,
\be
\Sigma_\pm=\frac{p_\pm}{-p_0}=\frac{d\b^\pm}{d\tau},\quad N_\a=\frac{12e^{2\b_\a}}{-p_0}.
\ee

In this setup of orthonormal, Hubble-normalized variables $\Sigma_\pm$ and $N_\a$, and Hamiltonian variables  $(\b^A,p_A)$, we introduce the length scale function $l$ by  setting $N=l^3=e^{\b_1+\b_2+\b_3},$ and $\tilde{t}:=T$ ($T$ is called the BKL time), so that $dt/dT=l^3$, and introduce the BKL variables as the state vector $(\b_\a,k_\a)$ in $\mathbb{R}^6$, with $d\b_\a/dT=k_\a$. It then follows that
\be
e^{\b_\a}=t^{p_\a},
\ee
where $p_\a$ are the Kasner exponents given by
\be
p_\a=\frac{k_\a}{k_1+k_2+k_3}.
\ee
This concludes the network of relations between the three sets of variables.

\subsection{Evolution equations and constraints for the spatial points}
We now return to our problem and consider two spatial `Mixmaster points' $A,B$ evolving towards the past singularity. We use for $A$ the standard normalized variables $X_A (t)=(N_1,N_2, N_3, \Sigma_+,\Sigma_-)$, as above,  and for $B$ a set of different such variables, $X_B (t)=(M_1,M_2, M_3, \Pi_+,\Pi_-)$, where the $N$'s and the  $M$'s depend on the Bianchi group structure and the trace part of the second fundamental form of the spacelike hypersurfaces, while the traceless part of the latter appears in the shear variables $\Sigma, \Pi$.

Then the variables  for the Mixmaster point $A$ evolve according to the vacuum Einstein equations \cite{em}, \cite{wh}, \cite{we},
\bq
N'_1&=& (q-4\Sigma_+)N_1,\label{n1}\\
N'_2 &=& (q+2\Sigma_+ + 2\sqrt{3}\Sigma_-)N_2, \label{n2}\\
N'_3 &=& (q+2\Sigma_+ - 2\sqrt{3}\Sigma_-)N_3,\label{n3} \\
\Sigma'_+ &=& -(2 - q)\Sigma_+ -3S_+, \label{+}\\
\Sigma'_-&=& -(2 - q)\Sigma_- - 3S_-,\label{-}
\eq
with the constraint,
\be\label{constraint}
\Sigma_+^2+\Sigma_-^2 +\frac{3}{2}\left(N_1^2+N_2^2+N_3^2 -2(N_1N_2+N_2N_3+N_3N_1)\right)=1,
\ee
a set which contains the Bianchi-IX attractor \cite{r1}. Here we have defined,
\bq
q&=&2(\Sigma_+^2+\Sigma_-^2),\\
S_+&=&\frac{1}{2}\left((N_2-N_3)^2-N_1(2N_1-N_2-N_3)\right),\label{splus}\\
S_-&=&\frac{\sqrt{3}}{2}(N_3-N_2)(N_1-N_2-N_3)\label{sminus},
\eq
and a prime denotes differentiation with respect to the $\tau$-time as above.

Similarly, for the second Mixmaster point $B$, which undergoes chaotic Mixmaster oscillations for the unknowns  $y=(M,\Pi)$,   the variables $M=(M_1,M_2,M_3)$ satisfy similar equations to the Eqns. (\ref{n1})-(\ref{n3}), and the shear variables $\Pi=(\Pi_+,\Pi_-)$ satisfy the system,
\bq
\Pi_{+}'&=&-(2-p)\Pi_+-3Q_+,\label{+pi}\\
\Pi_{-}'&=&-(2-p)\Pi_--3Q_-,\label{-pi}
\eq
with $p=2(\Pi_+^2+\Pi_-^2)$,  the $Q$'s like the $S$'s in the Eqns. (\ref{splus}), (\ref{sminus}) but with the $M$'s in the corresponding places of $N$'s, and the constraint identical to (\ref{constraint}), but with the $(M,\Pi)$'s in the places of the $(N,\Sigma)$'s.

We therefore end up with two systems, system-$A$ for the Mixmaster point $A$ evolving under the equations for the $(N,\Sigma)$ variables,  and system-$B$ evolving according to the corresponding equations for the $(M,\Pi)$ variables as above.

This concludes the equations to be satisfied during the evolution of two spatial points $A,B$ in an inhomogeneous Mixmaster spacetime. In the next Sections, we shall describe in detail the  interaction of these two points through the mechanism of dynamical synchronization.

\section{Generic synchronization}
\subsection{Sync in nonlinear dynamics}
Before turning to the problem of synchronized chaos in cosmology, we give a very brief description of the basic mechanism of sync  in general nonlinear dynamics. Synchronization of two chaotic systems  leading to  perfect harmony during their evolution was a totally unexpected feature in nonlinear dynamics when first discovered by Pecora and Carroll in their pioneering work \cite{pc} (see also \cite{fuji,afrai} for related earlier work,  \cite{pe97} for a review of their later work, \cite{ark} for a literature review with many references to various aspects of sync, and \cite{stro} for an elementary introduction). Pecora and Carroll had in mind applications to communications but also applied their method to the Lorenz and R\"{o}ssler systems.

One considers two copies of a chaotic system that play the roles of driver (transmitter) and receiver, the latter receiving signals from the driver, but does not send any back- it is a one-way communication.
To wit, suppose that one considers as the receiver a Lorenz system with variables $x,y,z$, and another copy of it, $x',y',z'$, a remotely evolving transmitter which sends a signal to the receiver replacing $x$ with $x'$ but leaving the other receiver variables $y,z$ unchanged. In \cite{pc}  it was  found that the receiver variables which were not replaced, namely  $y,z$, gradually would snap into sync with their counterparts $y',z'$ making all the variables of the receiver totally synchronized with the corresponding ones in the transmitter. Both systems after that were evolving in complete harmony.

What is remarkable about this result is that the receiver managed somehow to reconstruct the remaining  signal of the transmitter, that is the $y',z'$ part, by using only part of it-the variable $x'$. In some cases, it was later found that sync may in fact proceed exponentially rapidly \cite{cuomo93}. By using this method and after the work of \cite{pc} for the Lorenz and the R\"{o}ssler systems, many chaotic dynamical systems  have been found that allow for synchronization, a truly remarkable phenomenon that has since provided the key to many previously unresolved mysteries in nonlinear dynamics (for more  details, we refer to the  books \cite{mose}, \cite{wu}, and the recent review \cite{elp}).

\subsection{Sync in cosmology}
Because of relativistic causality, any two spatial points $A,B$ in an inhomogeneous cosmological spacetime may influence each other by sending signals in a one-way manner, $A$ can only influence $B$ if it lies in $B$'s causal past, and $B$ can only influence points on or inside its future null cone. Inspired by \cite{pc},
we may think of two Mixmaster points $A,B$ influencing each other as a `transmitter-receiver  pair'  (`master-slave', or `driving-response'), and examine the possibility of whether some form of sync may develop between them. We consider the evolution towards the past singularity, and imagine  that the evolving Mixmaster point $A$  touches $B$'s  past null cone and sends a signal to $B$. Then $B$ receives the signal of $A$ but does not send any back because $B$ is in $A$'s future. ($A$ may eventually stay inside the null cone of $B$ or leave-it is irrelevant.) In general, we may imagine that such a one-way causal communication constantly occurs between pairs of spatial points throughout some spacetime region.

Although the two spatial points in the pair both oscillate independently as two separate Mixmaster universes, we pose the question whether or not  the `receiver' $B$ achieves perfect synchronization with the transmitter point $A$ independently of any initial conditions, asymptotically towards the singularity gradually, and at an exponentially fast rate. In that case, $A$ and $B$ will thereafter oscillate  in absolute unison towards the singularity. In fact, the thought strikes one that  $B$ may be  able to do so even if it receives from $A$ only part of the information that fully characterizes the state of the transmitter point $A$, in a manner analogous to the Pecora-Carroll model described earlier. It is this problem a solution of which is found in this paper.

In the following we are not interested in the separate dynamics of two Mixmaster points $A,B$,  the first given in terms of the $(N,\Sigma)$ variables and the second in terms of  the $(M,\Pi)$ ones, but in their \emph{interaction dynamics}, their synchronization dynamics, which is a third, new,  dynamics satisfying equations to be derived below. We shall derive the fundamental equations for the synchronization dynamics of the two interacting Mixmaster points $A,B$ by taking the variables $N$ of the first point $A$ as the `driving' (or transmitter) variables, substitute the $N$ variables in the place of the $M$ variables of the second (receiver) point $B$ (as the result of communication between the two points), and study the evolution of the remaining  $\Pi=(\Pi_+,\Pi_-)$ of  $B$ regarded as the responding variables to the `driver' $N$.

In this approach, the expansion-normalized variables introduced in Section 2, which rewrite the problem as a dynamical system, have proven useful and this justifies our use of them in the present work. We think of the $A$-system in the variables $x=(N,\Sigma)$ as one of the form $x'=f(x)$, consisting of two subsystems with the $N$ variables satisfying the first three equations and the $\Sigma$'s the last two as above, namely,
\be
N'=g(N,\Sigma),\quad\Sigma'=h(N,\Sigma),
\ee
with $N=(N_1,N_2,N_3)$, $\Sigma=(\Sigma_+,\Sigma_-)$. We set the $B$-system (receiver) variables $M$ simply equal to the corresponding $A$-transmitter variables $N$, so that the $B$-system  equations become,
\bq
M_i&=&N_i,\quad i=1,2,3\label{13}\\
\Pi_{+}'&=&-(2-p)\Pi_+-3S_+,\label{+pi1}\\
\Pi_{-}'&=&-(2-p)\Pi_--3S_-.\label{-pi1}
\eq
We  introduce the \textit{synchronization function} $\Omega=(\Omega_+,\Omega_-)$, with $\Omega=\Sigma-\Pi$, that is:
\be\label{omega}
\Omega_+=\Sigma_+-\Pi_+,\quad \Omega_-=\Sigma_--\Pi_-.
\ee
We have complete synchronization of the two Mixmaster oscillating spatial points provided,
\be\label{synch}
\Omega\rightarrow (0,0),\quad \textrm{as}\quad \tau\rightarrow-\infty.
\ee
Otherwise, the two oscillating spatial points evolve autonomously, and stitching them together becomes impossible.

Using this geometric setup, what we show below is that even though the Mixmaster point $B$ receives only partial information, namely only the $N$ variables, about the state of the point $A$, it eventually manages to `reconstruct' the other two transmitter variables $\Sigma$ as well, and synchronize itself completely to $A$ on approach to the singularity; the two different spatial points are hereafter evolving in harmony as two indistinguishable homogeneous Mixmaster universes.

\subsection{Sync and Liapunov exponents}
Before we properly treat the sync problem in the next two Sections of this paper, we  present in the rest of this Section an argument which shows  that in the inhomogeneous  Mixmaster setup one in fact \emph{expects} some sort of generic synchronization to appear after some threshold value of the coupling between two spatial points is exceeded. We note that in general,  we expect  extra coupling terms in the different dynamical equations  of the two Mixmaster oscillating points taken together as prescribed by the dynamics of Eqns. (\ref{+}), (\ref{-}), and (\ref{+pi}), (\ref{-pi}).

The simplest non-trivial form of coupling is shown in the last terms of the next two equations, and is represented by  a coupling term \emph{linear} in the synchronization function $\Omega$,
\bq
\Sigma' &=& -(2 - q)\Sigma -3S+\a(\Pi-\Sigma)\label{s}\\
\Pi' &=& -(2 - p)\Pi -3Q+\a(\Sigma-\Pi),\label{p}
\eq
where $Q$ is the corresponding expression of $S$ for the $\Pi$ system, and the coupling terms here are taken to be proportional to $\Omega=\Sigma-\Pi$, with $\a$ the coupling constant. In general, the coupling terms will be described by a generic map,
\be
Z:\mathbb{R}^2\rightarrow\mathbb{R}^2:\Omega\mapsto Z(\Omega),
\ee
but here we simply choose $Z=I$, the identity map. 
Then from the synchronization system (\ref{s})-(\ref{p}), we can obtain sufficient conditions on the coupling constant $\a$ such that locally around the synchronization subspace $\Omega=0$, we have $\Omega(\tau)\rightarrow 0$ as the singularity is approached.

For the two Mixmaster trajectories of the points $A$ and $B$ described by the variables  $\Sigma,\Pi$ respectively, setting $Q=S$ (using Eq. (\ref{13})) and subtracting the equations  (\ref{s})-(\ref{p}), we find that,
\be
\Omega'=F(\Sigma)-F(\Pi)-2\a\Omega.
\ee
Hence, by Taylor expanding $F(\Omega)$ around $\Omega=0$ and dropping the second and higher-order terms, we obtain  the variational equation,
\be \label{var1}
\Omega'=DF(\Sigma-\a I)\Omega,
\ee
where $DF(\Sigma)$
is the Jacobian of $F$ evaluated at the Mixmaster trajectory $\Sigma$. Introducing  the variable
\be\label{omega1}
\omega(\tau)=e^{2\a\tau}\Omega,
\ee
we can rewrite Eq. (\ref{var1}) in the form,
\be
\omega' =DF(\Sigma)\omega,
\ee
which is the variational equation of the uncoupled Mixmaster system given by $\Sigma$.

Now it is a well-known result that if the orbit $\Sigma$ has Lyapunov exponent $\lambda$, then \cite{gh}
\be
|\omega|\leq Ce^{ \lambda \tau},
\ee
and so from Eq. (\ref{omega1}), we find
\be \label{aa}
|\Omega|\leq Ce^{(\lambda-2\a)\tau}.
\ee
Introducing the critical coupling strength
\be
\a_c=\frac{\lambda}{2},
\ee
we conclude that when
\be
\a>\a_c,
\ee
we have complete synchronization of the Mixmaster oscillating regions.

We know from old results about the Lyapunov exponents of Mixmaster dynamics that a common value is (see, eg., \cite{berger})
\be
\lambda\sim 0.45,
\ee
which in the present case  leads to the condition
\be \label{a1}
\a>0.225
\ee
for synchronization between the two spatial Mixmaster points $A,B$ in the original inhomogeneous spacetime. To compare, one requires $\a>0.4$ for two chaotic Lorenz oscillators, and $\a>0.03$ for two R\"{o}ssler systems, cf. \cite{elp}.

\section{Lyapunov function synchronization}
As we showed in the previous Section, the  synchronization map $Z$  introduces new coupling terms in the evolution equations of the oscillating spatial points $A$ and $B$ that  depend on $\Omega$, which in turn depends on the specific coupling of the oscillating Mixmaster points one uses.

However, as we now show, provided that the magnitudes-squared of the shear functions of the two spatial points are simply related, for instance, they are equal,
\be\label{shear}
p-q=0,
\ee
the system has a Lyapunov function which makes the state $\Omega=0$ globally asymptotically stable. In addition, sync between the two spatial points $A,B$ proceeds exponentially fast.

\subsection{The case of zero coupling}
First, suppose that there is no coupling between the two spatial points $A,B$. Then
under the assumption (\ref{shear}), the  equations (\ref{+}), (\ref{-}), and (\ref{+pi}), (\ref{-pi}) governing the dynamics of the problem, give synchronization equations of the form,
\bq
\Omega_{+}'&=&-(2-q)\Omega_+\label{seqns1}\\
\Omega_{-}'&=&-(2-q)\Omega_-,\label{seqns2}
\eq
since the terms involving $S,Q$ cancel out.  Multiplying the first equation by $\Omega_+$ and the second by  $\Omega_-$ and adding them together we find,
\be\label{oo}
\Omega_{+}\Omega_{+}'+\Omega_{-}\Omega_{-}'=-(2-q)(\Omega_+^2+\Omega_-^2).
\ee
We note that since the evolution is taking place inside the circle $q=2$, the term $2-q>0$. Further, since
\be
\Omega_{+}\Omega_{+}'+\Omega_{-}\Omega_{-}'=\frac{1}{2}\frac{d}{d\tau}\left(\Omega_+^2+\Omega_-^2\right),
\ee
we conclude that the function,
\be\label{pot}
V(\Omega_{+},\Omega_{-})=\frac{1}{2}\left(\Omega_+^2+\Omega_-^2\right),
\ee
satisfies $V(\Omega_{+},\Omega_{-})>0$, and its derivative is given by,
\be\label{dV}
\frac{dV}{d\tau}=-(2-q)(\Omega_+^2+\Omega_-^2).
\ee
Hence, it is negative everywhere inside the oscillating domain except at $\Omega=0$ where it vanishes. This result means that $V$ is a Lyapunov function for the synchronization dynamical equations (\ref{seqns1}),(\ref{seqns2}), and therefore the state $\Omega=0$ is globally asymptotically stable.

We conclude that the two oscillations points $A$ and $B$ must synchronize under the assumption (\ref{shear}). In fact,  using Eq. (\ref{dV}), we find that
\be\label{dV1}
\frac{dV}{d\tau}\leq-4V,
\ee
and so we arrive at an exponentially fast decay for $V$:
\be
V\leq V_0 e^{-4\tau},
\ee
where $V_0$ is a constant. Therefore all trajectories $\Omega(\tau)$ flow downhill toward the state $\Omega=0$ exponentially fast, and synchronization is stable asymptotically.

\subsection{More general shear-magnitudes relations}
We may somewhat extend this argument as follows. Suppose  that in place of the  assumption (\ref{shear}), we start by assuming  a more general form,
\be\label{a}
p=\alpha q,\quad \alpha \,\,\,\textrm{constant}.
\ee
Then introducing the functions
\be
\Omega^\alpha_{\pm}=\Sigma_{\pm}-\alpha\Pi_{\pm},
\ee
the synchronization equations (\ref{seqns1}),(\ref{seqns2}) become
\be
\Omega_{\pm}'=-2\Omega_{\pm}+q\Omega^\alpha_{\pm},
\ee
and so introducing the function $V$ as before, we see that its derivative satisfies,
\be\label{der1}
\frac{1}{2}\frac{dV}{d\tau}=-2(\Omega_+^2+\Omega_-^2)+q(\Omega_+\Omega^\alpha_{+}+\Omega_-\Omega^\alpha_{-}).
\ee
We can now show that this leads to another Lyapunov function for  synchronized dynamics. We first note that  the asymptotic order of the functions $\Omega^\a, \Omega$ is the same,
\be\label{o1}
\Omega^{\a}=O(\Omega),
\ee
meaning that there is a constant $K$ such that asymptotically towards the singularity, we have $|\Omega^{\a}|\leq K|\Omega|.$ To wit, since $O(\a\Pi)=O(\Pi)$, we find
\be
O(\Sigma-\Pi)=O(\textrm{max}(\Sigma,\Pi))=O(\textrm{max}(\Sigma,\a\Pi))=O(\Sigma-\a\Pi),
\ee
from which (\ref{o1}) follows. Therefore,
\be
O(\Omega\Omega^{\a})=O(\Omega)O(\Omega^{\a})=O(\Omega^2).
\ee
Using this result to rewrite the second term on the right side of Eq. (\ref{der1}), we find that asymptotically towards the singularity,
\be\label{der}
\frac{1}{2}\frac{dV}{d\tau}<-(2-q)(\Omega_+^2+\Omega_-^2)<0.
\ee
Therefore as before, $V$ is a Lyapunov function for the dynamics when the generalized condition  (\ref{a}) holds, and again synchronization proceeds exponentially fast.
The same argument may be further generalized if we assume
\be
p=f(\tau)q,
\ee
where $f(\tau)$ is a slowly varying function of the time.  This is then an example of generalized synchronization between the two oscillating points $A$ and $B$.

\subsection{The case of arbitrary coupling}
We now move on to consider the general case where the evolution equations contain not a coupling linear in the shear variables but a smooth function $f(\Omega)$ of them. The dynamical equations (\ref{+}), (\ref{-}), and (\ref{+pi}), (\ref{-pi}) are replaced by,
\bq
\Sigma' &=& -(2 - q)\Sigma -3S-f(\Omega)\label{s1}\\
\Pi' &=& -(2 - p)\Pi -3Q+f(\Omega),\label{p1}
\eq
leading to synchronization equations that replace Eqns.  (\ref{seqns1}), (\ref{seqns2}) to  include a general coupling term,
\bq
\Omega_{+}'&=&-(2-q)\Omega_+ -2f(\Omega)\label{seqns11}\\
\Omega_{-}'&=&-(2-q)\Omega_- -2f(\Omega)\label{seqns21}.
\eq
Following the procedure advanced earlier does not really work in this case because we need some general condition on $f$. We find that in the place of Eq. (\ref{oo}), we now have the equation,
\be\label{oo1}
\Omega_{+}\Omega_{+}'+\Omega_{-}\Omega_{-}'=-(2-q)(\Omega_+^2+\Omega_-^2)+g(\Omega),
\ee
with
\be
g(\Omega)=-2f(\Omega)(\Omega_+ +\Omega_-),
\ee
$g(\Omega)$ being a new term in the problem that is determined by the coupling function $f(\Omega)$.
Consider the function,
\be\label{pot1}
V(\Omega_{+},\Omega_{-})=\frac{1}{2}\left(\Omega_+^2+\Omega_-^2\right)+g(\Omega),
\ee
which we introduce as a possible Lyapunov function of the problem.

In order to check if this is indeed a Lyapunov function for the general dynamics, we assume  that $V\geq 0$, or,  equivalently,
\be\label{g}
g(\Omega)\geq -\frac{1}{2}\left(\Omega_+^2+\Omega_-^2\right),
\ee
and  calculate the derivative $dV/d\tau\equiv V'$ along orbits of the system (\ref{seqns11}), (\ref{seqns21}). This is given by,
\be
\frac{dV}{d\tau}=\Omega'\cdot\nabla V,
\ee
and  is sometimes called the orbital derivative \cite{wig}, with $\nabla V=(\pa V/\pa\Omega_+ ,\pa V/\pa\Omega_-)$. Using  this definition and the dynamical equations for this case, we find,
\be\label{dV11}
\frac{dV}{d\tau}=-(2-q)(\Omega_+^2+\Omega_-^2)+g(\Omega)+\Omega'\cdot\nabla g.
\ee
Therefore  provided that,
\be\label{dot}
\Omega'\cdot\nabla g \leq 0,
\ee
and also that $g$ is restricted to its negative values in the  range given in  (\ref{g}), we find that
$V$ is a Lyapunov function for this case and, hence, the state $\Omega=0$ is  asymptotically stable under these conditions. The assumption (\ref{dot}) is  a geometric condition on the vector field $\nabla g$ (and thus through $g$, on $f$), to be always either tangent or pointing inward for each of the level surfaces $g(\Omega)=C$ surrounding the state $\Omega =0$.

\section{Phase synchronization}
In the present Section, we consider the case where the two oscillating regions have their own different frequencies, leading to different dynamics for their phases. Under certain conditions, the two phases can synchronize while their amplitudes stay uncorrelated.
This   scheme is the situation encountered in the Barrow-Kuramoto model \cite{ba20}.

\subsection{Mean-field response of spatial points}
In the terminology of the present work, we consider $N$ Mixmaster spatial points, each one evolving according to the equations of Section 2, and interacting with each other with an equal coupling strength. The Mixmaster points have frequencies $\omega_i$ and phases $\theta_i\in [0,2\pi]$, normally distributed on $[0,2\pi)$. We define the mean-field created by the Mixmaster oscillating points by introducing a mean phase and its complex amplitude by the equation,
\be\label{r}
re^{i\psi}=\frac{1}{N}\sum_{j=1}^{N}e^{i\theta_j}.
\ee
Introducing polar coordinates (the arbitrariness of the polar radius $\rho$ is here reminiscent of the different sizes of the Kasner circles in different oscillating Mixmaster spatial points),
\be
\Sigma_+=\rho\cos\theta,\quad\Sigma_-=\rho\sin\theta,\label{po}
\ee
then for an arbitrary  Mixmaster oscillating point having phase $\theta$, we can  write down two remarkable equations describing how the oscillating point responds to the mean field quantities $r$ and $\psi$, \cite{ba20}:
\be\label{b1}
\theta'=\frac{3}{\rho}\left( S_+\sin\theta+S_-\cos\theta\right),
\ee
leading to
\be\label{b2}
\theta'=\frac{3}{\rho}\sqrt{S_+^2+S_-^2}\sin(\theta-\psi),\quad\tan\psi=S_-/S_+.
\ee

\subsection{Derivation of the phase-sync equations}
To facilitate the adaptation of phase-synchronization ideas in Mixmaster dynamics, we present an elementary proof of these two equations. Multiplying  the shear equations (\ref{+}), (\ref{-}) by $\sin\theta$ and $\cos\theta$ respectively, one has
\bq
\Sigma'_+\sin\theta &=& -(2 - q)\Sigma_+\sin\theta -3S_+\sin\theta, \label{++}\\
\Sigma'_-\cos\theta&=& -(2 - q)\Sigma_-\cos\theta - 3S_-\cos\theta.\label{--}
\eq
Subtracting the first from the second equation, using  Eq. (\ref{po}), and dividing by $\rho$, we obtain
\be
\frac{\Sigma'_-\cos\theta -\Sigma'_+\sin\theta}{\rho}=\frac{3}{\rho}(S_+\sin\theta -S_-\cos\theta).\label{pre}
\ee
Now expanding the left-hand-side of this equation using again Eq. (\ref{po}), one realizes that all terms miraculously cancel, leaving just $\theta'$ in that side of (\ref{pre}), which is then the sought-for equation (\ref{b1}). Further, using the trigonometric identity,
\be
a\sin\t +b\cos\t=\sqrt{a^2+b^2}\sin(\t +\delta),\quad\delta=\textrm{Arg}(a+ib),
\ee
and setting $a=S_+,b=S_-$, we find that the combination
\be
K=S_+\sin\theta+S_-\cos\theta,
\ee
in the right hand side of Eq. (\ref{b1}) can be rewritten in the form
\be
K=\sqrt{S^2_++S^2_-}\sin(\t +\delta),
\ee
where, the argument $\delta$ is given by,
\bq
\delta&=&\textrm{Arg}(a+ib)\\&=&\textrm{Arg}(S_+ -iS_-)\\&=&-\textrm{Arg}(S_+ +iS_-)\\&=&-\psi,\quad \psi=\textrm{Arg}(S_+ +iS_-),
\eq
or $\tan\psi=S_-/S_+$, as required. Then it follows that
\be
K=\sqrt{S^2_++S^2_-}\sin(\t -\psi),
\ee
which, using Eq. (\ref{b1}), leads to Eq. (\ref{b2}).

\subsection{Direct point-to-point couplings}
The fundamental relation  (\ref{b2}) dictates how any Mixmaster oscillating point responds to the mean field. However, under this equation, each spatial point appears to be totally independent of any other point, interacting only with the mean field. Of course this is not the case, and each Mixmaster oscillating point interacts with another one.

To see this explicitly, we multiply Eq. (\ref{r}) by $e^{-i\t_i}$, where $\t_i$ is the phase of the $i$-th oscillating region, to have
\be
re^{i(\psi-\t_i)}=\frac{1}{N}\sum_{j=1}^{N}e^{i(\theta_j-\t_i)},
\ee
and upon taking imaginary parts, we find that for the $i$-th region,
\be
r\sin(\psi-\t_i)=\frac{1}{N}\sum_{j=1}^{N}\sin(\theta_j-\t_i),
\ee
so that Eq. (\ref{b2}) gives,
\be\label{last}
\theta'_i=\frac{3}{N\rho_i}\sqrt{S_+^2+S_-^2}\sum_{j=1}^{N}\sin(\theta_j-\t_i),
\ee
i.e., we have the original Kuramoto model of spontaneous synchronization for directly coupled oscillators.

As noted in \cite{ba20},  the coupling in Eq. (\ref{b2}) is time-dependent, in distinction to the original model of Kuramoto. The same is of course true for the Eq. (\ref{last})  describing direct coupling of the Mixmaster points.

\section{Physical interpretation of Mixmaster sync}
\subsection{The problem}
The results of the previous Sections allow for an interesting physical interpretation and implication of synchronization in the present context of  spatial points in an inhomogeneous spacetime evolving as  separate homogeneous Mixmaster universes. Every spatial point is continually sending and receiving signals from other spacetime points, shifting its state of oscillations and adjusting it to that of other points, resetting their motion until sync organizes them in perfect harmony. How can this work in the present case? As in other manifestations of the mechanism of sync in nature \cite{stro}, we are led to suspect and hereby suggest  that each such spatial point is not only a Mixmaster oscillator, but a \emph{resettable} oscillator.

The problem may be phrased as follows. It is well-known (see eg., \cite{we}, chap. 11 and refs. therein) that the claimed chaotic behaviour of the Mixmaster dynamics as a process towards the past singularity involves different sequences of Kasner epochs and eras and is well described by iterates of the BKL map $f(u)$, in terms of the  BKL  parameter $u\in [1,\infty)$. The variable $u$  parameterizes the Kasner exponents $p_\a$, but in an inhomogeneous setup as  presently, one expects that $u$  takes different values in different spatial points. So following the evolution of two BKL maps at two different Mixmaster spatial points $A,B$, any two initial $u$-values will evolve after few steps into two uncorrelated sequences of their integer parts. Since the corresponding integer parts fix the number of Kasner epochs in each era,  the two sequences of epochs and eras of the two spatial points $A,B$ will be expected to be completely different. This may in fact lead to a remarkable form of `gravitational turbulence' \cite{bel} through the Kirillov-Kochnev-Montani effect mentioned in the Introduction which is based on the alternation of Kasner epochs.

\subsection{Synchronized BLK maps}
Then synchronization of  spatial points discussed previously may have a dramatic effect on this problem. Consider the two Kasner circles $\mathcal{K}^{\bigcirc}_A:\,\Sigma_+^2+\Sigma_-^2=1$, and  $\mathcal{K}^{\bigcirc}_B:\,\Pi_+^2+\Pi_-^2=1$, containing the Kasner equilibrium points of the dynamical systems (\ref{+}), (\ref{-}), and (\ref{+pi}), (\ref{-pi}), respectively. Since we have replaced the receiver  variables $(M_1,M_2,M_3)$ by the transmitter variables $(N_1,N_2,N_3)$, the circles $\mathcal{K}^{\bigcirc}_A, \mathcal{K}^{\bigcirc}_B$ contain all of the Kasner equilibria (which have $N_1=N_2=N_3=0$).

Then the Kasner exponents $p_\a^A,p_\a^B$ corresponding to the two evolving Mixmaster points $A,B$ are given by (cf. \cite{we}, Sec. 6.2.2, Eq. (6.16)),
\bq
p_1^A&=&\frac{1}{3}(1-2\Sigma_+),\quad p_{2,3}^A=\frac{1}{3}(1+\Sigma_+\pm\sqrt{3}\Sigma_-),\\
p_1^B&=&\frac{1}{3}(1-2\Pi_+),\quad p_{2,3}^B=\frac{1}{3}(1+\Pi_+\pm\sqrt{3}\Pi_-),
\eq
with $\sum p_\a^A=\sum (p_\a^A)^2=1$, and similarly for the $p_\a^B$'s.

But we have shown  that during the (exponentially fast) synchronization,  we have, $\Pi_{\pm}\rightarrow\Sigma_{\pm}$, because the sync function $\Omega\rightarrow 0$.

Therefore we find that,
\be\label{plim}
p_\a^B\rightarrow p_\a^A,\quad\a=1,2,3,\quad\textrm{as}\,\,\,\Omega\rightarrow 0,
\ee
and hence, the two sets of Kasner exponents become identical exponentially fast at the end of sync ($\Omega=0$).

We therefore  conclude that the two BKL parameters $u^A,u^B$,  parameterizing the two sets of Kasner exponents $p_\a^A,p_\a^B$ of the spatial points $A,B$ respectively (up to permutation symmetry), namely,
\be\label{pu}
p_1^{A,B}=\frac{-u^{A,B}}{1+u^{A,B}+(u^{A,B})^2},\quad p_2^{A,B}=\frac{1+u^{A,B}}{1+u^{A,B}+(u^{A,B})^2}, \quad
p_3^{A,B}=\frac{u^{A,B}(1+u^{A,B})}{1+u^{A,B}+(u^{A,B})^2},
\ee
will also converge,
\be
u^B\rightarrow u^A,
\ee
for $u^{A,B}\in (1,\infty)$, to the same value at the end of sync, for each one of the six sectors $(123), (132),\dots$, on the Kasner circles $\mathcal{K}^{\bigcirc}_A,\mathcal{K}^{\bigcirc}_B$, no matter the values they started with. They therefore give two sets of identical BKL map values after the end of sync at the two different spatial points.

The same is true for the boundary $u$-values of each sector, namely the values $u=1$,  and $u=\infty$ (this last value characterizes the so-called `Taub points' $T_1,T_2,T_3$, $p_\a=(1,0,0),(0,1,0),(0,0,1)$). In fact, the midpoints $Q_1,Q_2,Q_3$ of the three arcs connecting the three Taub points on a Kasner circle, which correspond to the LRS Kasner solutions, are also synchronized. This follows from their parametrization, cf. \cite{we}, p. 132, namely, by setting $\Sigma_+=\cos\psi,\Sigma_-=\sin\psi$, one finds that,
\be\label{psi}
p_1^{A,B}=\frac{1}{3}(1-2\cos\psi^{A,B}),\quad p_{2,3}^{A,B}=\frac{1}{3}(1+\cos\psi^{A,B}\pm\sqrt{3}\sin\psi^{A,B}),
\ee
with the $Q$ points lying at the values $\psi =0,2\pi/3,4\pi/3$, respectively. When $B$ receives data from $A$ and synchronizes,  then $\Pi_{\pm}\rightarrow\Sigma_{\pm}$, and so we have that $\psi^B\rightarrow\psi^A$, hence, $Q^B\rightarrow Q^A$. That is, the midpoints of the arcs connecting the Taub points on the two Kasner circles corresponding to $B, A$  also become indistinguishable. Hence, we conclude that the two $u$-values $u^A,u^B$ have been reset at the end of sync.

\subsection{Resetting effect and inversions}
To better understand the resetting effect in the $u^A,u^B$ values and the meaning of the phrase `at the end of sync' studied previously, we note the following circumstance with respect to out-of-sync inversions of the BKL maps associated with the two spatial points.

As we showed above, a state having
$|\Pi-\Sigma|\sim 0,$  so also $|\Omega|\sim0$, is reached exponentially fast during synchronization.  From Eqns. (\ref{plim}), (\ref{psi}) it follows that when the evolution reaches that  state, we will also have that,
\be
\label{psim0} |p_\a^A-p_\a^B|\sim 0.
\ee
Because of the continuity of the rational functions $p=p(u)$ given by Eqns. (\ref{pu}), we must also have,
\be\label{usync}
|u^A-u^B|\sim 0,
\ee
otherwise Eq. (\ref{psim0}), which holds because of the definition of limit (\ref{plim}), would be violated.

This implies that the sync between $A,B$ (that is, that the receiver point $B$ syncs with $A$) is in fact effected well \emph{before} the state $\Omega=0$ is reached. To see this, we note that during an era, starting from some initial value $u_1^B$, we obtain the sequence,
\be
u_1^B, u_1^B-1, u_1^B-2, \dots, u_1^B-n,
\ee
which proceeds until when $u_1^B-n=x_n^B$ has its integer part equal to zero  (similarly for the spatial point $A$). All such `end-of-an-era' $u$-values for the $A, B$ points are randomly distributed in the interval $(0,1)$. Then, any inversion
\be
x_n^B\rightarrow \frac{1}{x_n^B},
\ee
required for the $B$-evolution to continue past the $u_1^B$-era to the next one with starting value $u_2^B$,
\be
\frac{1}{x_n^B}\equiv u_2^B, u_2^B-1,\dots
\ee
can only occur in the present situation provided it does \emph{not} contradict Eq. (\ref{usync}) (and similarly for $A$). But such inversions will give nonzero integer parts to the sequences of the $u^A,u^B$, thus typically violating condition (\ref{usync}), the latter  however being valid  during sync as we proved earlier-a contradiction. Hence, sync-which as we have shown holds for the system of two spatial points, leads to the resetting of  their  BKL map sequences so that the two evolve in unison sharing the same BKL map.

We therefore find that the sync of the two spatial points $A,B$ is practically concluded during a few epochs once Eq. (\ref{usync}) is reached. The receiver point $B$  shifts its $u$-values suitably to become \emph{exactly equal} to those of $A$ during an era, because inversions creating values of $u^B$ having nonzero integer parts typically violate condition (\ref{usync}), and are therefore  forbidden soon after the process of sync has started. Hence, the spatial point $B$ will sync its $u$-values perfectly with those of $A$ during an era, according to condition (\ref{usync}).

\subsection{Physical meaning of Mixmaster sync}
This provides an important physical meaning of the mechanism of sync. The two Mixmaster points completely synchronize with one another  in the \emph{actual} sense: the numbers of Kasner epochs and eras for the  spatial points $A,B$ become identical exponentially fast during sync.

Therefore we conclude that the spatial point $B$ will synchronize with another point $A$ in its past, as above, as their sync function $\Omega\rightarrow 0$. This sync will become distinctively apparent when  their corresponding $u$-values become equal as shown in the previous Subsection, and the two points  start oscillating in perfect unison. As we have shown, this `resettable' aspect of the oscillators is exclusively an inhomogeneous effect, totally absent in the homogeneous Mixmaster evolution (where there is only one `spatial point'-the entire space).

This effect has another possible  implication. Although one does not have a natural scale when dealing with the vacuum Einstein equations as presently, if the two processes of creation of  different `cells' or `islands' in the Kirillov-Kochnev-Montani sense at the two spatial points $A, B$  become synchronized through the mechanism described above, then the fragmentation processes associated with the two spatial points $A,B$ described in Refs. \cite{kir1}-\cite{mon}, will also synchronize at the scale where sync has been completed between the two spatial points. This scale corresponds to the common $u$-value acquired by the two spatial points at the end of sync, as above.

Perhaps it would be an interesting problem to further examine the possibility of existence of a \emph{universal sync scale} responsible for controlling the fragmentation processes on the way to the singularity of a finite spacetime region containing synchronized spatial points. It is to be further noted that the effect of resetting the two oscillators to continue with a common $u$-value as shown above, implies the end of an otherwise expected unbounded growth in the spatial gradient of the metric towards the singularity between the two spatial points $A,B$.

\section{Concluding remarks}
In this paper we  introduced the possibility of chaotic synchronization of different spatial point trajectories as a new factor for inhomogeneous Mixmaster dynamics near the past singularity. We have demonstrated that if two  spatial points are coupled to each other  and one of them (`the transmitter')  communicates to the other (`the receiver') only part of the information necessary to determine its state, then the receiver will be able to reconstruct the remaining information and completely synchronize its chaotic evolution with that of the transmitter point.

Under certain conditions on the shear variables of the two spatial trajectories, we have established the existence of a Lyapunov function  leading to a globally asymptotically stable synchronized state such that the two spatial points evolve in complete and perfect harmony. This is apparent by consideration of their individual BKL maps, which accordingly become indistinguishable at the end of sync. This behaviour can be extended to any number of spatial points which start out of phase but gradually evolve  and  become perfectly synchronized, and this becomes apparent either directly or  through their response to the mean field created by an arbitrary number of Mixmaster oscillators.

We believe that this effect gives a clear  possible meaning of the phrase  `spatial points decouple towards the singularity' used in formulations of the BKL conjecture. It reinforces this picture in that it gives a \emph{sufficient} condition for `locality', the latter having the usual meaning that each spatial point's asymptotic evolution is described by an ordinary differential equation (as in \cite{u13} and refs. therein).

\subsection{Further issues}
In this work, we established the occurrence of sync between two spatial points evolving towards the singularity as two separate Mixmaster universes in an inhomogeneous generic spacetime.
We believe that our results are quite suggestive for a similar more general treatment of certain further points that were not considered presently.

The first issue is connected with the ability of signal reconstruction between the two spatial points: the spatial point $A$ in the past of the point  $B$ sends a signal to $B$ that contains the $N$-variable information of itself. Given this, $B$ is, as we showed in this work, then able to synchronize its shear variables to those of $A$. If the signal of $A$ contains other information of its state, say for instance only the $N_1$-variable, would $B$ then be able to synchronize like before and reconstruct the missing information? Indeed, what is the minimum amount of information transmitted  that the receiver could still use to reconstruct the state of the transmitter spatial point?

Secondly, the inclusion of matter is normally but conjecturally  taken not to influence the evolution of inhomogeneous Mixmaster spacetime towards the singularity. However,  there are no results that sync is also insensitive to the presence of matter. Although matter fields are insignificant towards the singularity, they could in principle   affect or possibly destroy the sync mechanism discussed here. More work on matter models is necessary before this issue is settled.

Further, our assumptions for the existence of a Lyapunov function ($p=q$, etc) for the sync dynamics are clearly restrictive. One may imagine that  two Mixmaster oscillating spatial points will also be able to synchronize not on the invariant set $\Sigma=\Pi$ but on some more general curve. Suppose that the two points have totally different vector fields $f_1, f_2$, but there is a general coupling between the `master' $\Sigma$ and `slave' $\Pi$ variables, $\Pi=\psi(\Sigma)$. If the trajectories  $\Sigma$  and $\Pi$  have the property that they tend to the function $\psi$ on approach to the initial singularity for some large coupling values, then the system will experience another kind of generalized synchronization. Otherwise there will be no such sync. This has been observed in the Lorenz and R\"{o}ssler systems, and  the same could in principle be applicable for the present problem. But there are no general conditions for this.

Also there are no results currently that support the conclusion that sync is a  typical feature  of other cosmologies. We do not know if  higher dimensionality affects synchronization dynamics, or whether or not there is some critical dimension after which chaos disappears in the `synchroverse' introduced here, like the dimension $d=10$ for the Mixmaster universe. Also we have no results about synchronization dynamics in extended theories of gravity or string cosmology, neither do we have any results about how the inclusion of a scalar field would affect this problem. Any results on this  perhaps simpler issue would be a first indication of sync in the non-chaotic, inhomogeneous  `asymptotically velocity-dominated' regime.

\subsection{Conjectural implications and future prospects}
Despite these obvious or almost obvious deficiencies of the present model, or perhaps because of a need to address them, we include in this final subsection a  discussion of certain conjectural  implications of cosmological sync.

First, one may ask whether the assumption (\ref{a}) is a valid one. One could use the present bounds on the shear from the temperature anisotropies of the CMB \cite{ba83}, \cite{ba85},
\be\label{pred}
|\Sigma|_0\sim1.6\times10^{-9},
\ee
to place very tight constraints on the $q=|\Sigma|^2$ (or $p$). However, reversing this argument, we may say that the primordial synchronization between a pair of oscillating regions $A, B$ as above, provides a  prediction for the present bounds on the shear. From the sufficient condition (\ref{a1}), we see that using (\ref{aa}),  any \textit{variation} in the shear to settle  to an acceptable present value requires only the consideration of a suitable number of Mixmaster oscillations in the future direction until today.

Secondly, the possible intricate relation between sync and the phenomenon of gravitationally induced turbulence mentioned earlier is an aspect of this work that clearly  needs to be analyzed further. With respect to the fragmentation process, the existence of a possible `universal' scale mentioned earlier where the process may stop over a large open set of synchronized regions, may lead to a completely new aspect of the inhomogeneous Mixmaster dynamics. If on the other hand, such a scale is not unique but sync is efficient over large distances of different patches, then we may end up with  well-defined, self-organized, `solitonic' regions which are synchronized and follow the BKL pattern of oscillations with some mild form of turbulence. This last possibility may relate to the case of `multi-fractal turbulence \cite{bafra}.

Thirdly,  quite apart from the above, consider the past evolution of two spatial points $D,E$ which lie on the same spacelike hypersurface $\mathcal{F}$ in inhomogeneous Mixmaster spacetime, and suppose that they are not in causal contact with each other. We consider two null cones with vertices $B,C$ that contain $D,E$ respectively (ie., there exist causal  paths from $D$ to $B$ and  from $E$ to $C$), and assume that their intersections  with $\mathcal{F}$ have no points in common. Using the sync mechanism suggested in this paper, $D$ will then synchronize with $B$, and $E$ with $C$. Now, consider the bigger null cone with vertex the point $A$  such that $B,C$  lie on it (i.e., both $B,C$ are on $A$'s past null cone). Since they are on $A$'s past, $B,C$ can both synchronize with $A$, which means that the original points $D,E$, although never in causal contact with each other, can acquire similar properties because they become in sync with the common point $A$. This argument suggests that sync may in fact lead to a new approach to the horizon problem.

Another possible implication of sync is related to the behaviour of entropy on approach to the past singularity. This is connected to the fact that each time two spatial point trajectories in inhomogeneous spacetime  synchronize with each other, the entropy of the system \emph{reduces} as a result of sync: information about the states  of the spatial points \emph{before} sync is lost after sync as they now evolve in unison to one another, so the asymptotic rate of creation of information towards the past singularity is thus reduced, cf. \cite{ru}. Therefore any region evolving near the past singularity and containing  synchronized spatial points possibly has lower entropy than what it would have had without sync. This  mechanism for entropy reduction does not  involve any `thermal' effects like those found in inflationary models. However, there are no estimates as to how efficient this process of entropy reduction can be, or how large  synchronized regions might become before hitting on the past singularity. This may also be related to the issue of whether or not sync has any connection to the issue of isotropization in the past, cf. \cite{lim}.

One last possibility related to the sync mechanism discussed here is about its connection to the silent singularity and the role of spikes and their oscillations \cite{uvwe,u13}. It seems that sync would be perfectly suited for tackling the silent singularity conjecture.  If we consider the past timelike trajectory of a spatial point, it is a standard expectation (cf. eg.,  \cite{euw}, Fig. 1) that the null cone of the vertex point will become more and more narrow and inhomogeneities will be pushed outside of them, hence residing outside regions of causal communications.  However, we have seen in the present paper that causally unconnected spatial points may in fact sync with each other, and this could possibly contribute in making the evolution along neighboring timelines decouple. Nevertheless, such decoupled spatial points will evolve in unison due to sync even if they were in causal communication only transiently or at all.

\section*{Acknowledgments}
This paper is dedicated to the fond memory of my mentor John David Barrow. John made pioneering contributions to the Mixmaster universe, and more generally to all aspects of cosmology, that extend to more than 40 years of extraordinary work. I am grateful to him for countless discussions, ideas, and  advice.

\end{document}